\documentclass[
nofootinbib,
 amsmath,amssymb,
 aps, 
 physrev, reprint
]{revtex4-2}

\usepackage[justification=raggedright,singlelinecheck=false]{caption}
\usepackage{subfig}
\usepackage{graphicx}
\usepackage{dcolumn}
\usepackage{bm}
\usepackage{xcolor}
\usepackage{color,soul}
\usepackage{hyperref}
\usepackage{mathtools}
\usepackage{braket}
\def\Ketbra#1#2{\left|#1\middle\rangle\middle\langle#2\right|}

\sethlcolor{yellow}
\soulregister\citet7

\begin{document}

\preprint{APS/123-QED}

\title{\textbf{Relativistic Maxwell-Bloch Equations with Applications to Astrophysics} 
}%

\author{Ningyan Fang}
\affiliation{%
  Department of Physics and Astronomy, The University of Western Ontario, 1151 Richmond Street, London, N6A 3K7, Ontario, Canada\relax
}%

\author{Victor Botez}
\affiliation{Laboratoire Interdisciplinaire des
Sciences du Numérique, CNRS, Campus universitaire Paris-Saclay, rue du
Belv\'ed\`ere, 91400 Orsay, France
}%

\author{Fereshteh Rajabi}
\affiliation{
  Department of Physics and Astronomy, McMaster University, 1280 Main Street West, Hamilton, L8S 4L8, Ontario, Canada
}%

\author{Martin Houde}%
\email{Contact author: mhoude2@uwo.ca}
\affiliation{%
  Department of Physics and Astronomy, The University of Western Ontario, 1151 Richmond Street, London, N6A 3K7, Ontario, Canada
}%
\date{\today}

\begin{abstract}
We derive relativistic Maxwell-Bloch equations for potential applications in astronomical environments, where various radiative processes are known to occur, including the maser action and Dicke's superradiance. We show that for both phenomena a radiating system's response is preserved at different relative velocities between the system's rest frame and the observer, while the relevant timescales and the radiation intensity transform as expected from relativistic considerations. We verify that the level of coherence between groups of emitters travelling at different speeds is unchanged in all reference frames. We also derive relativistic versions of the maser equations applicable in the steady-state regime.
\begin{description}
\item[Keywords]
relativity, coherence, superradiance, Maxwell-Bloch equations
\end{description}
\end{abstract}

\maketitle


\section{\label{sec:intro}Introduction}

The Maxwell-Bloch equations (MBEs) are a powerful tool for modelling light-matter interaction by tracking the time evolution of macroscopic parameters that characterize a system consisting of groups of molecules\footnote{We focus on molecules but our analysis applies to atoms, as well.} \citep{Bonifacio1965,Icsevgi1969,Lugiato2015}. With their wide range of applications in quantum optics, these equations are often used in the non-relativistic scheme, consistent with implementations in the laboratory setting \citep{McCall1967,Gross1982,Benedict1996}. While the MBEs are also relevant in several astronomical environments, where radiation processes are relativistic, however, they need to be modified to take relativistic effects into account.

Within this context, studying superradiance from relativistic sources provides a specific approach to a generalized version of the MBEs. First proposed by R. H. Dicke \citep{Dicke1954,Dicke1964}, superradiance has garnered increasing interest both theoretically and experimentally for several decades \citep{Gross1982,Benedict1996,Skribanowitz1973,MacGillivray1976,MacGillivray1981,Andreev1980,Arecchi1970}. It describes the coherent process of cooperative spontaneous emission emerging from a population of inverted molecules, which become coupled through their interaction with a common electromagnetic field and therefore must be treated as a single quantum mechanical system instead of independent radiators. Some main features of superradiance include enhanced radiation rate and shortened emission timescale. 

Recent works have provided observational evidence of the occurrence of Dicke's superradiance in astronomical media, specifically in maser hosting regions \citep{Rajabi2016b,Rajabi2017,Rajabi2019,Rajabi2020,Rajabi2023,Rashidi2025}. This is consistent with the fact that the maser action (stimulated emission) and superradiance are complementary phenomena occurring in different regimes of the MBEs \citep{Feld1980, Rajabi2020}. 

It has furthermore been proposed that superradiance could be at the origin of fast radio bursts (FRBs) \citep{Houde2018,Houde2019}. As this requires the existence of sources moving at relativistic speeds relative to the observer, a triggered relativistic dynamical model (TRDM), motivated by superradiance, was introduced to model FRB emission \citep{Rajabi2020a}. It consists of a trigger source located in the background behind the FRB source, which moves at a relativistic velocity and emits highly collimated radiation towards an observer in the foreground. The TRDM predicts an inverse relationship between the slope of the frequency of an FRB sub-burst's intensity and its duration (as seen in a dynamic spectrum showing the the radiation intensity as a function of frequency and time). This relationship, known as the sub-burst slope law, has since been validated with multiple sources and data sets \citep{Rajabi2020a,Chamma2021,Jahns2023,Chamma2023,Brown2024,Chamma2025}.  

Motivated by such studies, our interest resides in investigating superradiance and the maser action in the relativistic scenario. This problem was first addressed, in part, by considering superradiance from a relativistic source for the simple case of two two-level particles in the Schr\"{o}dinger picture \citep{Wyenberg2022}. Yet, we would like to find a more generalized solution for larger systems. To achieve this, we tackle the problem by considering MBEs for relativistic molecular ensembles. We look to answer the following questions: if it is known that superradiance takes place in the rest frame, what would an observer see as a result? How do the timescales and intensities transform? What role does velocity coherence (see Sec.~\ref{sec:results} for a definition) play in this process? 

We start in Sec.~\ref{sec:theory} by introducing the theoretical framework. We derive relativistic MBEs, for the case where the radiation propagates in a direction parallel to the molecules' velocity, based on a fully relativistic Hamiltonian in the observer's frame and demonstrate the timescale transformation in the linear regime. We then use the same model to derive relativistic maser equations. With the methodology introduced in Sec.~\ref{sec:method}, the numerical results and discussion are presented in Sec.~\ref{sec:results} and shown to be consistent with the prediction of the TRDM. We re-examine the requirement for velocity coherence in the relativistic context, followed by a short conclusion in Sec.~\ref{sec:conclusion}. Finally, a more detailed derivation of the Hamiltonian can be found in Appendix \ref{sec:derivation}.

\section{Theory}\label{sec:theory}

\subsection{Derivation of the relativistic MBEs}\label{subsec:MBEs}
We consider an ensemble of $N$ initially excited, identical molecules of mass $m$ approximated as two-level systems, which move collectively at the same relativistic velocity towards or away from an observer, while interacting with the ambient electromagnetic field. We define the primed quantities in the rest frame and those in the observer's frame unprimed. The Hamiltonian in position space in the observer's frame reads

\begin{align}
\hat{H}&=\hat{H}_a+\hat{H}_f+\hat{H}_{\text{int}}\label{eq:total Hamiltonian}
\end{align}
with
\begin{align}
\hat{H}_a&=\sum_{j=1}^N\left(\gamma mc^2\Ketbra{\textbf{r}_j}{\textbf{r}_j}+\frac{\hbar\omega_0'}{\gamma}\hat{R}_3^{\prime j}\right)\label{eq:atomic Hamiltonian}\\
\hat{H}_f&=\sum_{p}\hbar\omega_{p}\hat{a}_p^{\dagger}\hat{a}_p\label{eq:field Hamiltonian}\\
\hat{H}_\text{int}&=-\sum_{j=1}^N\sum_p\left[\left(\textbf{d}'\cdot\bm{\epsilon}_p\right)\left(\hat{R}_+^{\prime j}+\hat{R}_-^{\prime j}\right)\left(\hat{E}_p^++\hat{E}_p^-\right)\right],\label{eq:interaction Hamiltonian}
\end{align}
where $\hat{H}_a$, $\hat{H}_f$, and $\hat{H}_\text{int}$ represent the molecular, free radiation field (neglecting the vacuum energy), and the electric dipole interaction Hamiltonians, respectively. The latter applies when photons propagate in a direction parallel (or anti-parallel) to the molecules' velocity, which is expected for relativistic superradiance from the large, elongated systems considered in Sec.~\ref{sec:results} (see Appendix~\ref{sec:derivation}). The mixed nature of the Hamiltonian, with all internal molecular quantities defined in the source's rest frame and the radiation field pertaining to the observer's, allow us to use parameters where they are more naturally defined. For example, the electric dipole moment $d^\prime$ is only truly known (and measured) in the rest frame of the emitter while the quantity we are most interested in, i.e., the radiation field, is that measured by the observer. The mechanical and internal energies of the molecules in $\hat{H}_a$ transform relativistically with the Lorentz factor $\gamma=\left(1-\beta^2\right)^{-1/2}$ for $\beta=v_0/c$ the relative velocity to the observer. For the $j^\mathrm{th}$ molecule, whose upper and lower states are $|b\rangle$ and $|a\rangle$, $\hat{R}_3^{\prime j}=\frac{1}{2}\left(|b\rangle\langle b|-|a\rangle\langle a|\right)$. The Bohr frequency in the rest frame $\omega_0'$ is associated with the transition with an electric dipole moment $\textbf{d}'$ of magnitude $d'$ and orientation $\bm{\epsilon}_d'=\textbf{d}'/d'$. It is assumed that $\bf{d}'$ is the same for all molecules in the system. The creation and annihilation operators of photons of radiation mode $p$ and linear polarization $\bm{\epsilon}_p$ are $\hat{a}_p^\dagger$ and $\hat{a}_p$, respectively. The internal energy state of the molecules is raised and lowered by $\hat{R}^{\prime j}_+=|b\rangle\langle a|$ and $\hat{R}^{\prime j}_-=|a\rangle\langle b|$. Together with $\hat{R}^{\prime j}_3$, these operators are analogous to those used in the spin-1/2 formalism. Note that we use the long-wavelength approximation where the molecules' position $\textbf{r}_j$ is treated classically \citep{Grynberg2010}. The quantized electric field operator is 
\begin{align}
    \hat{\textbf{E}}^+&=i\sum_p\hat{E}_p\bm{\epsilon}_p\nonumber\\
    &=i\sum_p\mathcal{E}_p\hat{a}_pe^{i\textbf{k}_p\cdot\textbf{r}}\bm{\epsilon}_p \label{eq:field operator}
\end{align}
with the single-photon electric field $\mathcal{E}_p$ and wave vector $\mathbf{k}_p$ (also $\hat{\mathbf{E}}^-=\left.\hat{\mathbf{E}}^+\right.^\dagger$). The detailed derivation of the Hamiltonian is provided in Appendix \ref{sec:derivation}.

To obtain the relativistic MBEs, we define (half of) the population density difference between the upper and lower levels and polarization in the rest frame using the observer's frame coordinates as
\begin{align}
    \hat{n}'(\textbf{r})&=\frac{1}{\gamma}\sum_{j=1}^N\delta\left(\textbf{r}-\textbf{r}_j\right)\hat{R}'^j_3\label{eq:population density}\\
    \hat{\textbf{P}}'^\pm(\textbf{r})&=\frac{1}{\gamma}\,\textbf{d}'\sum_{j=1}^N\delta\left(\textbf{r}-\textbf{r}_j\right)\hat{R}'^j_\pm.\label{eq:polarization primed}
\end{align}

With equations (\ref{eq:field operator})-(\ref{eq:polarization primed}), as well as the Heisenberg equation for the time evolution of operators, the fully relativistic one-dimensional MBEs are found to be (with the photon energy much lower than $\gamma mc^2$; see Appendix~\ref{sec:derivation})
\begin{align}
    \frac{d\hat{n}'_v}{d\tau}&=-\frac{1}{i\hbar}\left(\hat{P}_v^{\prime+}\hat{E}^+-\hat{P}_v^{\prime-}\hat{E}^-\right)-\frac{\hat{n}_v'}{T_1}+\hat{\Lambda}_{n'_v}\label{eq:1stMBE}\\
    \frac{d\hat{P}^{\prime+}_v}{d\tau}&=ik\Delta v\hat{P}^{\prime+}_{v}+\frac{2id'^{2}}{\hbar}\hat{n}'_v\hat{E}^--\frac{\hat{P}_v^{\prime+}}{T_2}+\hat{\Lambda}_{P'_v}\label{eq:2ndMBE}\\
    \frac{\partial\hat{E}^+}{\partial z}&=\gamma\sqrt{\frac{1-\beta}{1+\beta}}\frac{i\omega_0'}{2c\epsilon_0}\sum_v\hat{P}^{\prime-}_v\label{eq:3rdMBE},
\end{align}
where the subscript $v=v_0+\Delta v$ denotes the velocity at which the groups of molecules are moving with respect to the observer, with $\Delta v$ the offset from central velocity $v_0=\beta c$. This velocity offset is related to that in the source's rest frame through $\Delta v \simeq \Delta v^\prime\left(1-\beta^2\right)$, valid when $\Delta v^\prime\ll c$. The temporal derivatives are relative to the retarded time $\tau=t-z/c$. The population density difference between the two levels is sustained by a pump at rate $\hat{\Lambda}_{n'_v}$, while $\hat{\Lambda}_{P'_v}$ stands for the initial vacuum fluctuations in polarization that initiate the response of the system. Relaxation and dephasing processes are phenomenologically introduced through the (observer's frame) timescales $T_1$ and $T_2$, which account for the depletion of population density and the decay of coherence, respectively. As implied by the spatial derivative in equation (\ref{eq:3rdMBE}), the radiation from the system is assumed longitudinal along the $z$-axis. Accordingly, the slow-varying envelope approximation (SVEA) is applied in the derivation of equations (\ref{eq:1stMBE})-(\ref{eq:3rdMBE}), where the electric field and the polarization consist of a fast-oscillating exponential and slow-varying envelope operators:
\begin{align}
    \hat{\textbf{E}}^\pm(\textbf{r},t)&=\hat{E}^\pm(z,\tau)\,e^{\mp i\omega\tau}\boldsymbol{\epsilon}_d\label{eq:SVEA_E}\\
    \hat{\textbf{P}}_v^{\prime\pm}(\textbf{r},t)&=\hat{P}_v^{\prime\pm}(z,\tau)\,e^{\pm i\omega\tau}\boldsymbol{\epsilon}'_d,\label{eq:SVEA_P'}
\end{align}
where we set $\boldsymbol{\epsilon}_d=\boldsymbol{\epsilon}_d'$ for simplicity. The envelopes $\hat{E}^\pm$ and $\hat{P}_v^{\prime\pm}$ vary over temporal and spatial scales much longer than $\omega^{-1}$ and $k^{-1}$ (with $k=\omega/c$), respectively. The radiation frequency in the observer's frame $\omega$ is related to the Bohr frequency (in the rest frame) by the Doppler relation
\begin{align}
    \omega=\gamma\omega_0'\left(1+\beta\right)\label{eq:Doppler}.
\end{align}

Equations (\ref{eq:1stMBE})-(\ref{eq:3rdMBE}) are the final form of MBEs that we will use for the numerical analyses in the following sections. Being one-dimensional in nature, the relativistic MBEs lend themselves well to the study of the propagation or evolution of an electromagnetic field of the type given in equation (\ref{eq:SVEA_E}) within a gas. Furthermore, because we expect coherence, and that coherence in radiation cannot exist over an arbitrary large volume, an MBE analysis is then equivalent to restricting ourselves to a long cylindrical system constrained to a Fresnel number of unity, i.e., $F=A'/\lambda'L'=1$, with $\lambda'$ the radiation wavelength, and $L'$ and $A'$ the length and cross-section of the cylinder, respectively. The outgoing radiation field will then be limited to an extremely small solid angle $\sim \lambda'/L'$ by such geometry, since $L'\gg \sqrt{A'}\gg\lambda'$, which allows us to focus on coherent emission along the $z$-axis and neglect off-axis spontaneous emission \citep{Gross1982}.  

It is also worth highlighting the differences between the standard and relativistic MBEs here. While the quantities in standard MBEs are all defined in the same reference frame, in relativistic MBEs, as previously stated, the population density difference and polarization are rest frame quantities, whereas the electric field is measured in the observer's frame. The timescales, as seen by the observer, are Lorentz transformed from their rest frame counterparts (see Sec.~\ref{sec:linear} below). 
The population inversion and polarization pumps are a hybrid of quantities from both the rest and observer's frames. Their detailed definition will be found in Sec.~\ref{sec:results}. Finally, the most prominent relativistic correction is reflected by the factor of $\gamma\sqrt{\left(1-\beta\right)/\left(1+\beta\right)}$ in equation (\ref{eq:3rdMBE}). 

From now on, the operator ``hat'' notation (e.g., $\hat{E}_p$) will be omitted as we apply a mean-field approach and replace the operators by their classical counterparts \citep{Gross1982}. We further remove the subscript $v$ since we assume that uniform pumps $\Lambda_{n'}$ and $\Lambda_{P'}$ are applied to all channels. Unless otherwise specified, we consider the velocity channel at central velocity $v_0$, i.e., $\Delta v=0$.

\subsection{Linear Regime}\label{sec:linear}

We consider the linear regime where the radiation intensity and polarization are weak. It is then safe to assume that $dn'/d\tau\simeq0$ since the second-order terms $P^{\prime\pm}E^{\prime\pm}$ become negligible in equation (\ref{eq:1stMBE}), whose solution is found to be $n'=\Lambda_{n^\prime}T_1\equiv n'_0$. In the fast transient limit, $\partial P^{\prime+}/\partial\tau\gg P^{\prime+}/T_2$, one can write
\begin{align}
    \frac{\partial P^{\prime+}}{\partial\tau}&=\frac{2id'^2}{\hbar}E^-n'_0\label{eq:dP/dtau}\\
    \frac{\partial E^+}{\partial z}&=\gamma\frac{i\omega}{2c\epsilon_0}P^{\prime-}.\label{eq:dE/dz}
\end{align}
These two equations can be combined to yield
\begin{align}
    \frac{\partial^2P^{\prime+}}{\partial\tau\partial z}&=\frac{1}{L'T_R'}\frac{1}{1-\beta}P^{\prime+}\nonumber\\
    &=\frac{1}{LT_R}P^{\prime+},\label{eq:linear regime}
\end{align}
where the characteristic superradiance timescale \citep{MacGillivray1976,Gross1982} in the rest frame is
\begin{align}
    T_R'&=\frac{8\pi}{3\lambda^{\prime2}n'_{\text{t}}L'\Gamma'}\label{eq:T_R'}
\end{align}
with $\lambda'$ the radiation wavelength, the total population density difference $n'_{\text{t}}=2n_0'$, $L'$ the sample length, and $\Gamma'$ the spontaneous emission rate, all defined in the rest frame. With length contraction $L'=\gamma L$, the characteristic superradiance timescale in the observer's frame is expressed as
\begin{equation}
    T_R=\sqrt{\frac{1-\beta}{1+\beta}}T_R'.\label{eq:T_R}
\end{equation}
This result is consistent with the relativistic transformation of time intervals. The relaxation and dephasing timescales that appear in equations (\ref{eq:1stMBE}) and (\ref{eq:2ndMBE}) thus follow the same transformation, i.e., $T_1=\sqrt{\left(1-\beta\right)/\left(1+\beta\right)}\,T_1'$ and $\,T_2=\sqrt{\left(1-\beta\right)/\left(1+\beta\right)}\,T_2'$.

\subsection{Relativistic maser equations}

We can derive relativistic maser equations using our MBEs, as well. From \citet{Feld1980} and \citet{Rajabi2020}, we know that while superradiance occurs in the transient regime, the maser action corresponds to the quasi-steady state limit when $\partial n'/\partial\tau\simeq\partial P'/\partial\tau\simeq 0$. The first two MBEs then transform to
\begin{align}
    n'&=\frac{n_0'}{1+I/I_{\text{sat}}}\label{eq:n'}\\
    P^{\prime+}&=\frac{2id'^2T_2}{\hbar}n'E^-\label{eq:maser polarization},
\end{align}
where the radiation and saturation intensities are, respectively, expressed as
\begin{align}
    I&=\frac{c\epsilon_0}{2}\left|E^+\right|^2\label{eq:I}\\
    I_{\text{sat}}&=\frac{c\epsilon_0\hbar^2}{8d'^2T_1T_2}.\label{eq:saturation intensity}
\end{align}
The ratio $I/I_\text{sat}$ in equation (\ref{eq:n'}) is an invariant and thus can also be expressed as $I'/I'_\text{sat}$, with $I'$ and $I'_\text{sat}$ defined similarly in the source's rest frame. Inserting equation (\ref{eq:maser polarization}) into the last of the MBEs, one can write
\begin{equation}
    \frac{dI}{dz}=\alpha I,\label{eq:dI/dz}
\end{equation}
where the gain coefficient in the observer's frame is
\begin{align}
    \alpha&=\gamma\frac{2\omega_0'd'^2T_2'n'}{c\epsilon_0\hbar}\nonumber\\
    &=\gamma\alpha'\label{eq:gain}
\end{align}
with $\alpha'$ the gain coefficient in the rest frame. Alternatively, we can write
\begin{align}
    \alpha'=\frac{2T_2'}{L'T_R'}\label{eq:alpha'}
\end{align}
and similarly for $\alpha$ in the observer's frame. 

In the weak field limit, $I\ll I_{\text{sat}},\alpha$ is a constant since $n^\prime=n^\prime_0$ from equation (\ref{eq:n'}). Solving equation (\ref{eq:dI/dz}) gives
\begin{align}
    I(z)&=I_0e^{\alpha z}\nonumber\\
    &=\left(\frac{1+\beta}{1-\beta}\right)I_0'e^{\alpha'z'}\label{eq:unsaturated maser intensity}
\end{align}
with $I_0$ and $I_0'$ are the background intensity at the input of the system at $z=z'=0$, respectively, in the observer's and rest frames. The two intensities are related by an amplifying factor of $\left(1+\beta\right)/\left(1-\beta\right)$ due to the Lorentz transformation of the electric field, whereas the exponential is invariant resulting from equation (\ref{eq:gain}) and length contraction $z=z'/\gamma$. The radiation intensity in both frames increases exponentially with distance, which is a salient feature of unsaturated masers.

On the other hand, in the strong field limit $I\gg I_{\text{sat}}$, $\alpha$ is inversely proportional to $I$ from equation (\ref{eq:n'}),
which leads to a linear solution
\begin{align}
    I(z)&=\left(\frac{1+\beta}{1-\beta}\right)\frac{\hbar\omega_0'n'_{\text{t}}}{8T_1'}\,\gamma z\nonumber\\
    &=\left(\frac{1+\beta}{1-\beta}\right)I'(z').\label{eq:saturated_maser}
\end{align}
The relativistic corrections are thus manifested by the appearance of the same factor $\left(1+\beta\right)/\left(1-\beta\right)$ in the last equation.

\section{Methodology}\label{sec:method}

The relativistic Maxwell-Bloch equations are numerically solved using a fourth-order Runge-Kutta method \citep{Houde2019}. As previously stated, the timescales $T_1$ and $T_2$ are added to our MBEs in order to account for relevant relaxation and dephasing processes that respectively affect the population density difference and polarization. Being  phenomenological in nature, they are not directly derived from the Hamiltonian alone. The same applies to the inversion $\Lambda_{n'}$ and polarization $\Lambda_{P'}$ pumps. While the nature of $\Lambda_{n'}$ will be detailed in Sec.~\ref{sec:results}, the polarization pump is set to $\Lambda_{P'}=n_0'd'\sin{\theta_0}/T_2$, with $\theta_0=2/\sqrt{N}$ the initial tipping (Bloch) angle characterizing the quantum fluctuations that initiate the response of the system at $\tau=0$ \citep{Gross1982}. Finally, in accordance with the retarded time framework, the two pumps propagate longitudinally along the sample ($+z$-direction). 

\section{Results and Discussion}\label{sec:results}

\begin{figure}
    \centering
    \includegraphics[width=\linewidth]{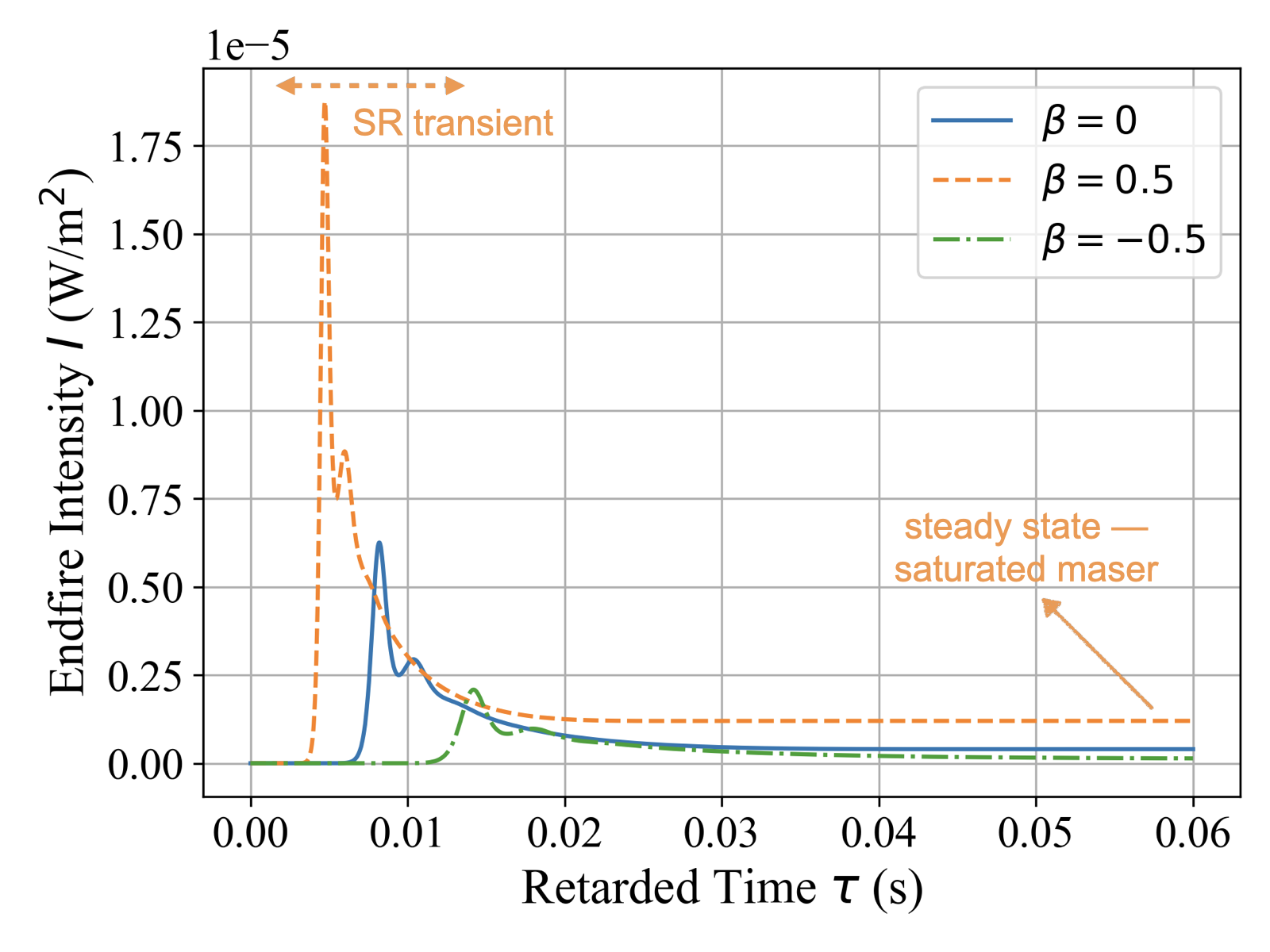}
    \caption{Superradiance (SR) and maser emission at $z=L'/\gamma$ for relative velocities $\beta=0$ (solid blue), 0.5 (dashed orange), and $-0.5$ (dash-dotted green) with $n_t'=2\times10^{4}$ m$^{-3}$. The shape of the curves is preserved regardless of the velocity but squeezed/stretched depending on the direction of motion relative to the observer.}
    \label{fig:1}
\end{figure}

Bearing in mind that the following quantities are defined in the rest frame of the molecules, we study the 1612 MHz spectral transition of the OH molecule ($\lambda'=18.6$~cm and $\Gamma'=1.282\times10^{-11}~\text{s}^{-1}$ for the Einstein coefficient of spontaneous emission) and base our parameters on those used in \citet{Houde2019}, where superradiance in this spectral line was investigated as a tentative model for FRBs. Accordingly, we set a length $L'=4.2\times10^{13}$~m (280~au) and a population density difference $n_\mathrm{t}'=2\times 10^{4}$~m$^{-3}$. The latter corresponds to an inversion level on the order of $1\%$ for a total gas density of approximately $10^7\,\mathrm{cm^{-3}}$ spread over $\sim 1\,\mathrm{km\,s^{-1}}$ for a hydroxyl abundance of $10^{-6}$. The total number of molecules partaking in the radiation process within the spectral extent of the signal thus amounts to $N\simeq 7\times 10^{30}$. While the population inversion fraction (1\%) corresponds to an approximate upper limit as suggested in the astronomical maser literature \citep{Elitzur1992}, it is not critical since the important parameter directly related to superradiance is the inverted column density. As seen in equation (\ref{eq:T_R'}), the superradiance characteristic timescale varies inversely with the inverted column density, i.e., $T_R'\propto 1/n_\mathrm{t}'L'$. It follows that a change in the population density difference $n_\mathrm{t}^\prime$ can be compensated by an opposite change in the length of the sample.

In order for superradiance to ensue, the dephasing timescale must be greater than the characteristic timescale of superradiance, which is evaluated to $T_R'=22\,\mu\mathrm{s}$ using the above parameters and equation (\ref{eq:T_R'}). We therefore set $T_2'=1.2\,\mathrm{ms}$ and assign a fiducial value of $T_1'=0.1$~s for the relaxation timescale, which is consistent with the fact $T_1'>T_2'$, generally. This value for $T_1'$ was chosen such that the (constant) inversion pump $\Lambda_{n'}=n_0'/T_1$ is strong enough to sustain a measurable saturated maser regime in the steady-state. The pump $\Lambda_{n'}$ acts as an effective pump that phenomenologically accounts for the different group of molecular transitions and interactions with the surroundings that lead to a net change in the population density of the two-level system \citep{Rajabi2023}. For hydroxyl, it is believed that far-infrared radiation at 35 $\mu$m and 53 $\mu$m is responsible for the population inversion \citep{Rajabi2016b,Gray2012}. As previously mentioned, the polarization pump $\Lambda_{P'}=n_0'd'\sin{\theta_0}/T_2$, with $\theta_0=2/\sqrt{N}$. We focus on the ``endfire" radiation mode, in which highly collimated radiation along the $+z$-direction is emitted from the corresponding end of the sample, as seen by an observer located at $z=L$. It is assumed that the direction of both the radiation and motion of the molecules are aligned along the line of sight to the observer, as this is also the condition under which we derived our relativistic MBEs.

Fig.~\ref{fig:1} shows the endfire radiation intensities of a sample of length $L=L'/\gamma$ for various velocities of the molecules relative to the observer. The positive or negative sign of $\beta$ denotes the direction of motion, i.e., $\beta=0.5$ represents the molecules moving relativistically towards the observer at half the speed of light, while $\beta=-0.5$ represents the molecules receding away from the observer at the same speed. We observe that significant superradiant response, which is characterized by a time delay preceding the arrival of the burst of radiation, occurs in the transient regime in all three scenarios before eventually transitioning to saturated maser emission in the steady state. As stated above, this constant, non-vanishing intensity towards the end of the simulation is sustained by the constant population inversion pump $\Lambda_{n'}$. As can be inferred from the common intensity profile, the system appears to the observer as essentially responding the same way regardless of the relative velocity. The duration of the pulse is, however, compressed when $\beta>0$ and stretched when $\beta<0$ by a factor of $\sqrt{\left(1-\beta\right)/\left(1+\beta\right)}$, as expected from the relativistic transformation of timescales (see equation \ref{eq:T_R}). On the other hand, the radiation intensity is modified by a factor of $\left(1+\beta\right)/\left(1-\beta\right)$ due to the Lorentz transformation of the electric field. As can be seen in Fig.~\ref{fig:1}, the peak intensity for $\beta=0.5$ is about three times of that for $\beta=0$, and the latter is about three times of the peak intensity at $\beta=-0.5$. The same is true for the steady-state intensities, as was already established with equation (\ref{eq:saturated_maser}). This result is in agreement with the predictions of the TRDM model for FRBs, based on relativistic transformations alone. Also, we note that the relativistic beaming effect, which is unaccounted for in the MBEs, would modify the radiation intensities by an additional factor of $\left(1+\beta\right)/\left(1-\beta\right)$ \citep{Rajabi2020a}.

\begin{figure}
    \centering
    \includegraphics[width=\linewidth]{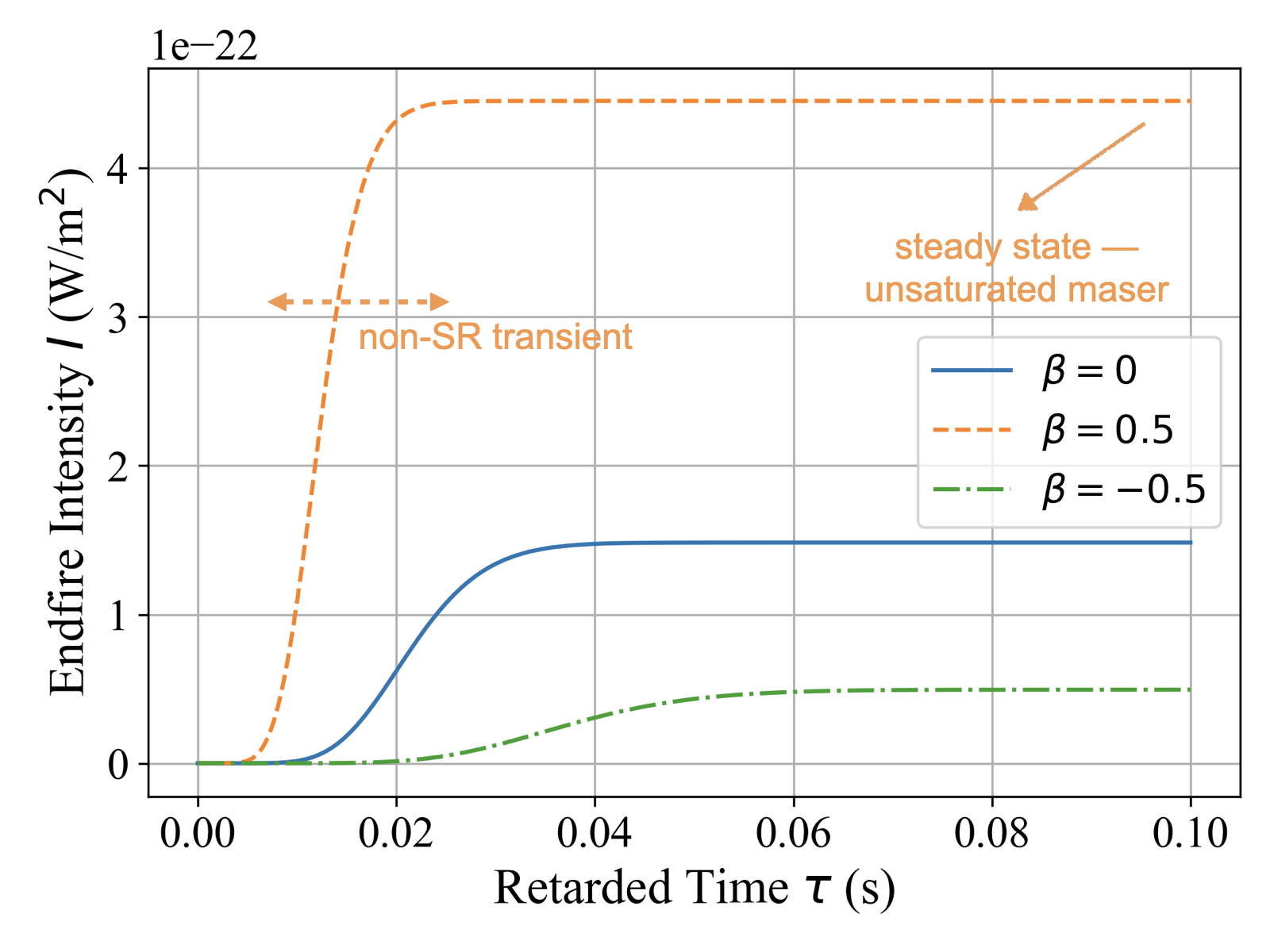}
    \caption{Same as Fig.~\ref{fig:1} with $n_t'=6\times10^{3}$ m$^{-3}$. Similar observations can be made for the intensities and timescales as for Fig.~\ref{fig:1}.}
    \label{fig:2}
\end{figure}

Fig.~\ref{fig:2} shows emission for the same system moving at various relative velocities with a reduced population density difference $n'_t=6\times10^3\,\text{m}^{-3}$. This inversion level is below the critical threshold needed for a superradiance response in the transient regime (see \citet{Rajabi2020}). Accordingly, we do not observe any overshoot or ringing as in Fig.~\ref{fig:1} and, therefore, the system exhibits a non-superradiance (``non-SR'' in the figure) response in the transient regime. The comparatively weak level of radiation in the steady-state is a characteristic of the unsaturated maser regime. Note that a decrease by a factor of $\sim 3$ in the initial population density difference resulted in a reduction of almost 16 orders of magnitude in radiation intensity. Otherwise, we reach the same conclusion as from Fig.~\ref{fig:1}-- the features of the system's response are preserved, apart from the relativistic corrections to intensity and timescales. 

So far, we have been concerned with situations where all molecules are moving collectively at the same relativistic velocity relative to the observer. This automatically fulfills the requirement of velocity coherence, which allows the molecules to interact through their common radiation field. This constraint can be expressed as $\left(\textbf{e}_q\cdot\Delta\textbf{v}\right)/c\leq\Delta\nu/\nu$, with $\textbf{e}_q$ the unit vector along the direction of emission, $\Delta\textbf{v}$ the approximate required velocity difference between emitters, $\nu$ the radiation frequency, and $\Delta\nu\sim1/\Delta t$ the frequency extent of the signal and $\Delta t$ its duration. Any difference in velocity between groups of molecules would therefore impede their cooperation and reduce the superradiant response. The separation between these channels is then crucial to the behaviour observed. 

Assuming the velocity channels follow a narrow, uniform distribution, one can define a fundamental velocity step in the rest frame $dv'=\left(2\pi c/\omega'_0\right)\tau^{-1}_{\text{max}}$ as well as a given velocity separation between the channels $\Delta v'=kdv'$, with $\tau_{\text{max}}$ the maximum simulation (retarded) time and $k$ a positive integer. Equivalently, the velocity separation can be expressed as a frequency difference $\Delta\omega'=kd\omega'$ as $d\omega'=\omega_0'dv'/c$. As previously mentioned, the velocity separation in the rest frame is related to that in the observer's frame through $\Delta v \simeq \Delta v^\prime\left(1-\beta^2\right)$ (with $\Delta v^\prime\ll c$). As previously stated, our model assumes that molecular movements are one-dimensional, therefore only the two directions parallel or anti-parallel to the radiation emission are considered, while velocity channels denote the actual velocity at which certain groups of molecules are correspondingly moving in the rest frame of the source. This assumption is consistent with the fact superradiance is a highly collimated type of radiation (because of the inherent phase coherence, which can only be maintained over a small solid angle \mbox{\cite{Gross1982}}), while relativistic motions will further amplify this beaming effect \mbox{\cite{Rybicki1979,Jackson1999}}.


\begin{figure}
    \centering
    \includegraphics[width=\linewidth]{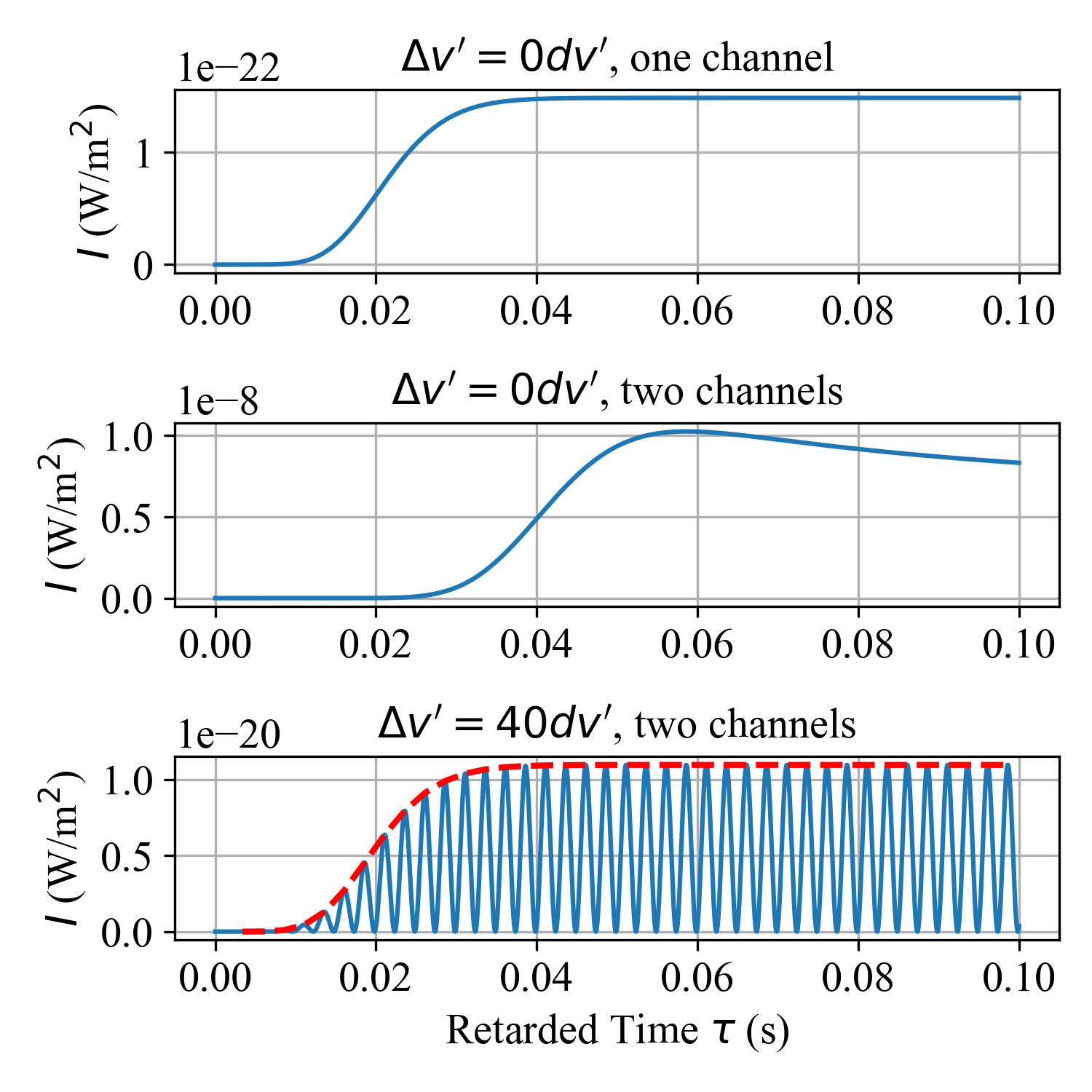}
    \caption{Endfire intensities of systems in the rest frame with one or two velocity channels, each channel of population density difference $n'_t=6\times10^{3}$\,m$^{-3}$. From top to bottom, the panels correspond to systems of non-superradiant single channel, superradiant two-channel with $\Delta v'=0$, and non-superradiant two-channel with $\Delta v'=40\,dv'$, respectively.}
    \label{fig:3}
\end{figure}

We will first investigate the impact of a separation between two velocity channels in the rest frame ($\beta = 0$). As shown in Fig.~\ref{fig:3}, in the top panel we have a system of single velocity channel at resonance of population density difference $n'_t=6\times10^{3}$ m$^{-3}$. No superradiance is observed in this case, but as we introduce a second velocity channel with the same $n_t'$ and $\Delta v'=0$ (middle panel), the system exhibits superradiance features in the transient regime: after a delay time of about 0.02 s, an overshoot of intensity occurs. Given that the two channels are perfectly aligned in velocity, this is equivalent to a system with a single channel of $2n_t'=1.2\times10^4$ m$^{-3}$, which (when multiplied by $L'$) exceeds the column density threshold for superradiance to take place. If the two velocity channels are instead separated by $40\,dv'$ centred around $v'_0=0$, as shown in the bottom panel, the system's response again converges to that of a single velocity channel, highlighted by the red broken envelope. The superradiant features seen in the middle panel vanish as a result of the weak coupling between the channels, and the group of molecules within each channel evolves more or less independently. Although the velocity separation subdued superradiance, the remaining weak coupling still produces an intensity almost two orders of magnitude stronger than in the top panel. The fast oscillations due to beating between the frequencies of the two channels are clearly observed. 

\begin{figure}
    \centering
    \includegraphics[width=\linewidth]{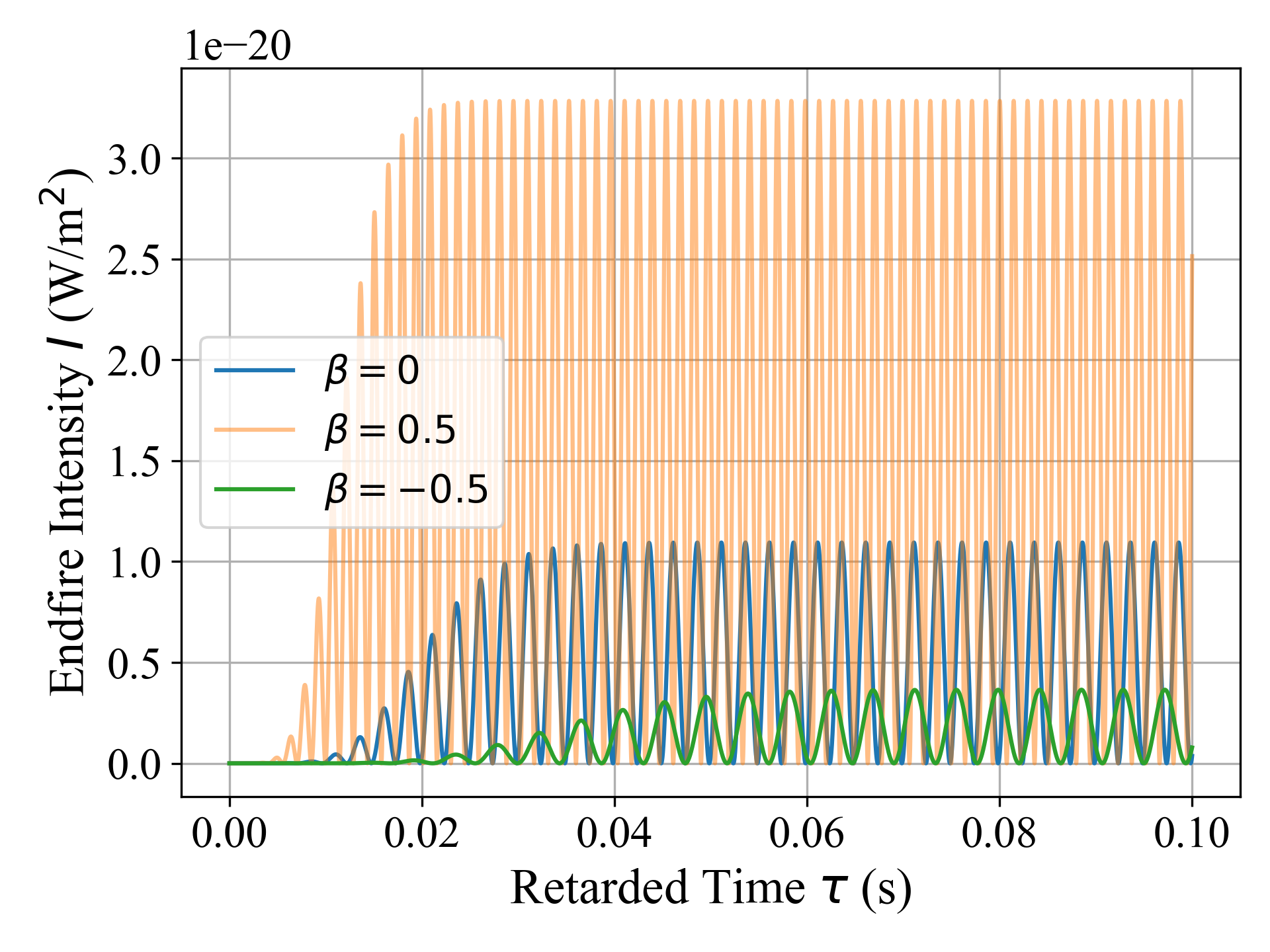}
    \caption{Endfire intensities of a two-channel system with $\Delta v'=40\,dv'$ moving at centre velocities $\beta=0$ (blue), 0.5 (orange) and $-0.5$ (green) relative to the observer. Both channels possess a rest frame population density difference of $n'_t=6\times10^{3}$\,m$^{-3}$. As expected, the same level of interaction is maintained, while the intensities scale with the factor $(1+\beta)/(1-\beta)$.}
    \label{fig:4}
\end{figure}

We now examine the same two-channel system with $\Delta v'=40\,dv'$ moving at relativistic velocities relative to the observer. Intuitively, if the separation between the two channels is such that they do not interact in the rest frame, they should also not appear to interact in the observer's frame either. Fig.~\ref{fig:4} shows the endfire intensities of the aforementioned system moving at $\beta=0$ (identical to the bottom panel in Fig.~\ref{fig:3}) and $\beta=\pm 0.5$. The level of coherence between the two channels is preserved regardless of the relative velocity, which can be inferred from the unchanged intensity profile (i.e., the envelope). This is because the ratio of the separation in frequency between the two channels and the spectral extent of the signal is a relativistic invariant since both quantities transform according to the Doppler shift law. The level of coupling between the channels is thus also a relativistic invariant. Otherwise, we also note that the radiation intensities follow the same scaling factor $\left(1+\beta\right)/\left(1-\beta\right)$ as before.

\begin{figure}
    \centering
    \includegraphics[width=\linewidth]{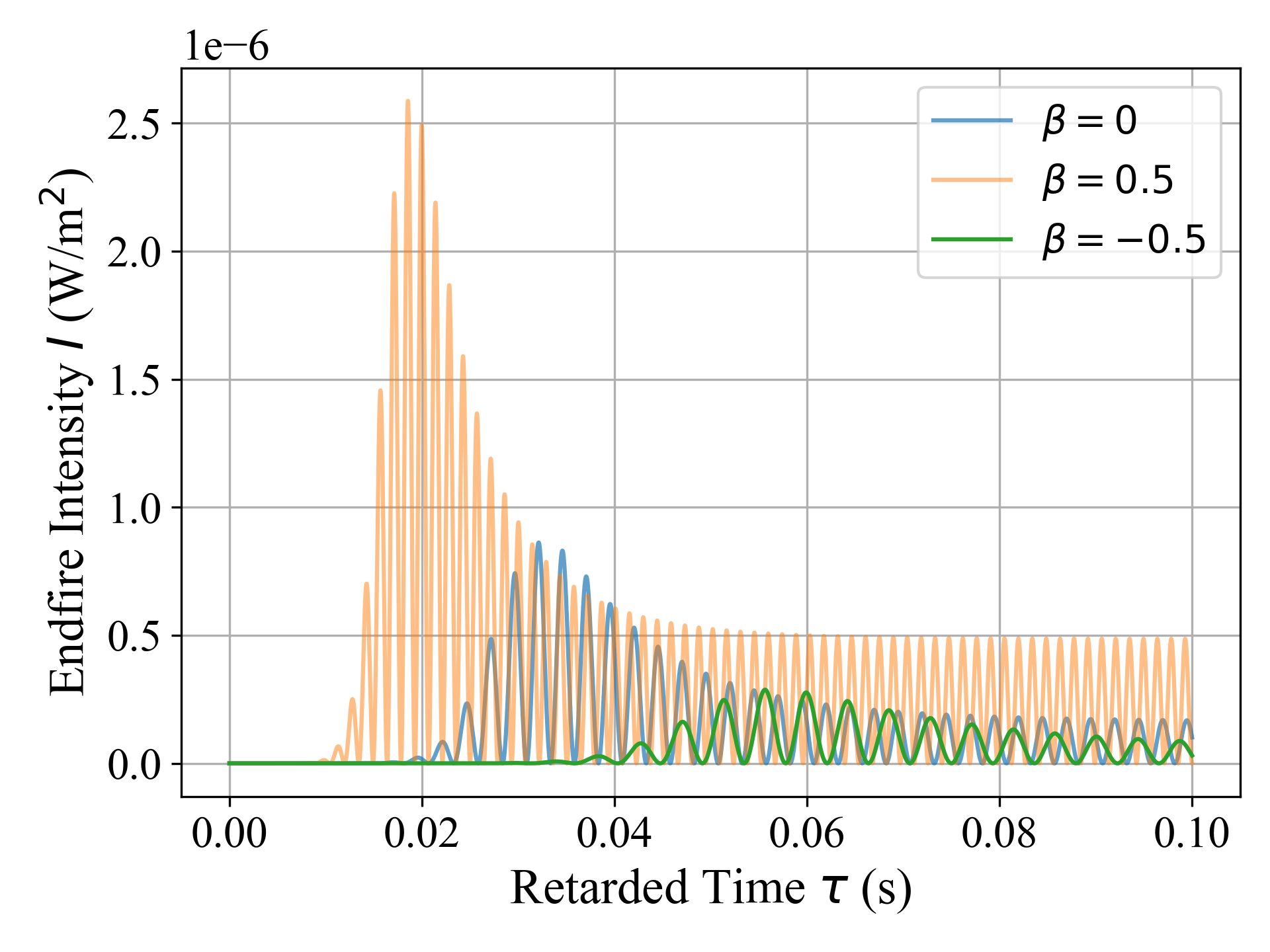}
    \caption{Same as Figure~\ref{fig:4} but with the population density difference of each channel increased to $n_t'=1.2\times10^4\,\text{m}^{-3}$. A superradiance response is observed at all three velocities, with the same intensity scaling factor described above.}
    \label{fig:5}
\end{figure}

Similarly in Fig.~\ref{fig:5}, when setting the population density difference of each channel to $n_t'=1.2\times10^4$ m$^{-3}$, the decoupling stemming from the velocity separation is overcome and the system shows superradiance again. The salient superradiant features that occur in the rest frame carry on to the relativistic scenarios as well. The same observations concerning the beating frequency and peak intensities can be made. As expected, the requirement for velocity coherence, and its level, does not change from one reference frame to another.

Finally, we would like to discuss potential applications of our theoretical formalism. The relativistic MBEs derived here are not restricted to astrophysical environments but apply generally to relativistically moving quantum emitters possessing internal transitions, such as molecules or ions. Laboratory realizations of such systems exist in laser spectroscopy experiments on relativistic ion beams in storage rings, where internal transitions are driven and fluorescence is observed while the emitters travel at relativistic velocities (e.g., \citep{Schroder1990,Saathoff2003,Gassner2018,Winzen2021}). These experiments confirm that light–matter interaction involving internal quantum transitions remains well defined under relativistic motion, with frequencies and timescales transforming according to relativistic principles, consistent with the framework developed here. While current experiments typically probe non-inverted systems, the formalism presented in this work provides a theoretical basis for describing coherent amplification and collective radiative phenomena in relativistically moving ensembles, should inversion be achieved through appropriate pumping schemes with sufficient column densities.


\section{Conclusion}\label{sec:conclusion}

We derived relativistic MBEs and used them to model superradiance and maser emission in the transient and steady-state regimes, respectively. Our analysis reveals that while the system's response is preserved at different relative velocities, the changes in intensity and timescales agree with predictions from relativistic transformations alone. The requirement for velocity coherence, which is essential for superradiance to occur, and the coupling between molecules moving at different velocities are maintained in any reference frame. Our MBEs can be applied to a broader range of phenomena and used to model various relativistic radiative processes.

\begin{acknowledgments}
M.H.’s research is funded through the Natural Sciences and Engineering Research Council of Canada (NSERC) Discovery Grant RGPIN-2024-05242 and the Western Strategic Support for Research Accelerator Success. F.R.’s research is supported by the NSERC Discovery Grant RGPIN-2024-06346.
\end{acknowledgments}

\appendix

\section{Derivation of the relativistic Hamiltonian}\label{sec:derivation}

We adopt the formalism developed by \citet{Boussiakou2002} that addresses the relativistic corrections of the Hamiltonian of a system composed of a two-level atom and the background radiation field using the Power-Zienau-Woolley gauge \citep{Tannoudji1989,Power1959,Woolley1971}. Since the Hamiltonian is not a Lorentz invariant, we start with the Lagrangian density for the electromagnetic field and the matter-field interaction
\begin{align}
    \mathcal{L}=-\frac{\epsilon_0}{4}F^{\mu\nu}F_{\mu\nu}-\frac{1}{2}D^{\mu\nu}F_{\mu\nu},\label{eq:Lagrangian}
\end{align}
where $F^{\mu\nu}$ is the electromagnetic tensor and $D^{\mu\nu}$ the polarization field tensor. The tensor $D^{\mu\nu}$ has the same form as $F^{\mu\nu}$ but with the substitutions $\textbf{E}\rightarrow \textbf{P}$ and $c\textbf{B}\rightarrow -\textbf{M}/c$, with \textbf{P} and \textbf{M} the polarization and magnetization, respectively. Hence \textbf{P} and $-\textbf{M}/c$ Lorentz transform the same way as \textbf{E} and $c\textbf{B}$:
\begin{align}
    \textbf{P}&=\textbf{P}'_\parallel\left(\textbf{r}'\right)+\gamma\left[\textbf{P}'_\perp\left(\textbf{r}'\right)+\frac{1}{c^2}\dot{\textbf{R}}\times\textbf{M}'\left(\textbf{r}'\right)\right]\label{eq:polarization}\\
    \textbf{M}&=\textbf{M}'_\parallel\left(\textbf{r}'\right)+\gamma\left[\textbf{M}'_\perp\left(\textbf{r}'\right)-\dot{\textbf{R}}\times\textbf{P}'\left(\textbf{r}'\right)\right]\label{eq:magnetization}
\end{align}
with $\dot{\textbf{R}}$ the centre of mass velocity of the molecule. The parallel and perpendicular subscripts are with respect to $\dot{\textbf{R}}$. Since we only consider the electric dipole interaction, $\textbf{M}'$ is zero and equations (\ref{eq:polarization})-(\ref{eq:magnetization}) allow us to write the interaction Lagrangian in the observer's frame as follows
\begin{align}
    L_{\text{int}}&=\int d^3r\left\{\left[\textbf{P}'_\parallel\left(\textbf{r}'\right)+\gamma\textbf{P}'_\perp\left(\textbf{r}'\right)\right]\cdot\textbf{E}\left(\textbf{r}\right)\right.\nonumber\\
    &\quad\qquad\left.-\gamma\left[\dot{\textbf{R}}\times\textbf{P}'\left(\textbf{r}'\right)\right]\cdot\textbf{B}\left(\textbf{r}\right)\right\}.\label{eq:interaction Lagrangian}
\end{align}

If we move on to the quantum mechanical description and neglect the vacuum energy, the Hamiltonian in the observer's frame reads
\begin{align}
    \hat{H}&=\hat{H}_a+\hat{H}_f+\hat{H}_{\text{int}}\label{eq:total Hamiltonian (BBB)}
\end{align}
with
\begin{align}
    \hat{H}_a&=\gamma mc^2\left|\textbf{r}\rangle\langle\textbf{r}\right|+\frac{\hbar\omega_0'}{\gamma}\hat{R}_3^{\prime}\label{eq:atomic Hamiltonian (BBB)}\\
    \hat{H}_f&=\sum_p\hbar\omega_p\hat{a}_p^\dagger\hat{a}_p\label{eq:field Hamiltonian (BBB)}\\
    \hat{H}_{\text{int}}&=\int d^3r \Bigg\{\frac{1}{\epsilon_0}\left[\hat{\textbf{P}}'_{\parallel}\left(\textbf{r}'\right)+\gamma\hat{\textbf{P}}'_{\perp}\left(\textbf{r}'\right)\right]\cdot\hat{\boldsymbol{\Pi}}\left(\textbf{r}\right)\nonumber\\
    &+\frac{\hat{\bf{p}}}{2m}\cdot\left[\hat{\bf{P}}'\left(\bf{r}'\right)\times\hat{\bf{B}}\left(\bf{r}\right)\right]+\left[\hat{\bf{P}}'\left(\bf{r}'\right)\times\hat{\bf{B}}\left(\bf{r}\right)\right]\cdot\frac{\hat{\bf{p}}}{2m}\Bigg\}.\label{eq:interaction Hamiltonian (BBB)}
\end{align}
We remind the readers again that the primed quantities are defined in the rest frame and the unprimed observer's frame. The centre of mass linear momentum is denoted by $\hat{\bf{p}}=m\dot{\textbf{R}}$, and $\hat{\bf{\Pi}}\left(\bf{r}\right)=-\epsilon_0\hat{\bf{E}}\left(\bf{r}\right)$. The terms that involve the magnetic field in the observer's frame $\hat{\bf{B}}\left(\bf{r}\right)$ are the so-called R$\ddot{\text{o}}$ntgen term, stemming from symmetrization necessitated by the non-commuting nature of position and linear momentum.  If we define 
\begin{align}
\hat{\bf{P}}'\left(\bf{r}'\right)&=\hat{\bf{P}}^{\prime -}\left(\bf{r}'\right)+\hat{\bf{P}}^{\prime +}\left(\bf{r}'\right)\nonumber\\
&=\textbf{d}'\sum_i\left(\hat{R}_-^{\prime i}+\hat{R}_+^{\prime i}\right)\delta\left(\textbf{r}'-\textbf{r}'_i\right),\label{eq:polarization definition}
\end{align} 
upon using the Lorentz transformation linking the coordinates of the two frames, we find
\begin{align}
    \hat{\bf{P}}'\left(\bf{r}\right)=\frac{1}{\gamma}\textbf{d}'\sum_i\left(\hat{R}_-^{\prime i}+\hat{R}_+^{\prime i}\right)\delta\left(\textbf{r}-\textbf{r}_i\right).\label{eq:polarization (BBB)}
\end{align}
Performing the integral in equation (\ref{eq:interaction Hamiltonian (BBB)}) yields
\begin{align}
    \hat{H}_\text{int}&=\frac{1}{\epsilon_0}\left(\frac{1}{\gamma}\hat{\textbf{d}}'_\parallel+\hat{\textbf{d}}'_\perp\right)\cdot\hat{\bf{\Pi}}\nonumber\\
    &\quad+\frac{\hat{\bf{p}}}{2\gamma m}\cdot\left(\hat{\bf{d}}'\times\hat{\bf{B}}\right)+\left(\hat{\bf{d}}'\times\hat{\bf{B}}\right)\cdot\frac{\hat{\bf{p}}}{2\gamma m}.\label{eq:new interaction Hamiltonian}
\end{align}

Expanding the Hamiltonian to a system of $N$ identical molecules, all at resonance with the radiation field, we have
\begin{align}
    \hat{H}&=\sum_{j=1}^N\left(\gamma mc^2\Ketbra{\textbf{r}_j}{\textbf{r}_j}+\frac{\hbar\omega_0'}{\gamma}\hat{R}_3^{\prime j}\right)+\sum_p\hbar\omega_p\hat{a}_p^\dagger\hat{a}_p\nonumber\\
    &-\sum_{j=1}^N\sum_p\left\{\left[\left(\frac{1}{\gamma}\hat{\textbf{d}}'_\parallel+\hat{\textbf{d}}'_\perp\right)\cdot\boldsymbol{\epsilon}_p\right]\right.\nonumber\\
    &\qquad\qquad\cdot\left(\hat{R}_-^{\prime j}+\hat{R}_+^{\prime j}\right)\left(\hat{E}_p^++\hat{E}_p^-\right)\nonumber\\
    &+\frac{1}{2\gamma mc}\left[\hat{\textbf{p}}_j\cdot\left(\textbf{d}'\times\boldsymbol{\epsilon}_p^\perp\right)+\left(\textbf{d}'\times\boldsymbol{\epsilon}_p^\perp\right)\cdot\hat{\textbf{p}}_j\right]\nonumber\\
    &\qquad\qquad\left.\rule{0mm}{5mm}\cdot\left(\hat{R}_-^{\prime j}+\hat{R}_+^{\prime j}\right)\left(\hat{E}_p^++\hat{E}_p^-\right)\right\}\label{eq:revised Hamiltonian}
\end{align}
where $\hat{\textbf{d}}'=\hat{\textbf{d}}'_\parallel+\hat{\textbf{d}}'_\perp$ with $\bf{d}'_\parallel$ and $\bf{d}'_\perp$ oriented parallel and perpendicular to the velocity (and the momentum $\bf{p}$) of the molecules. We use $\hat{\textbf{B}}_p=\boldsymbol{\epsilon}^\perp_p \hat{E}_p/c$ with $\boldsymbol{\epsilon}^\perp_p$ the polarization mode perpendicular to that of the electric field $\boldsymbol{\epsilon}_p$, and the electric field $\hat{E}_p$ defined in equation (\ref{eq:field operator}).  Both polarization modes are perpendicular to the direction of the wave vector $\bf{k}$, such that $\textbf{k},\,\boldsymbol{\epsilon}_p,\,\text{and}\,\boldsymbol{\epsilon}_p^\perp$ form a triad. As explained in Sec.~\ref{sec:results}, superradiance produces highly collimated radiation, which is further focused through relativistic beaming, implying that photons travel along the same or opposite direction as the molecules, namely, $\bf{p}\parallel\bf{k}$, we then have $\textbf{d}'_\parallel\cdot\boldsymbol{\epsilon}_p=\textbf{d}'_\parallel\cdot\boldsymbol{\epsilon}_p^\perp=0$ and thus $\left(\frac{1}{\gamma}\hat{\textbf{d}}'_\parallel+\hat{\textbf{d}}'_\perp\right)\cdot\boldsymbol{\epsilon}_p=\textbf{d}'\cdot\boldsymbol{\epsilon}_p$. 

Upon using the long-wavelength approximation and treating the molecules' coordinates as classical quantities \citep{Grynberg2010}, the last term in the Hamiltonian in equation (\ref{eq:revised Hamiltonian}) can be further simplified to
\begin{align}
    &\frac{1}{2\gamma mc}\left[\hat{\textbf{p}}_j\cdot\left(\textbf{d}'\times\boldsymbol{\epsilon}_p^\perp\right)+\left(\textbf{d}'\times\boldsymbol{\epsilon}_p^\perp\right)\cdot\hat{\textbf{p}}_j\right]\nonumber\\
    &\quad\cdot\left(\hat{R}_-^{\prime j}+\hat{R}_+^{\prime j}\right)\left(\hat{E}_p^++\hat{E}_p^-\right) =\nonumber\\
    &\qquad\qquad\frac{\hbar\omega}{\gamma mc^2}\left(\textbf{d}'\cdot\boldsymbol{\epsilon}_p\right)\left(\hat{R}_-^{\prime j}+\hat{R}_+^{\prime j}\right)\left(\hat{E}_p^+-\hat{E}_p^-\right)
\end{align}
using $\hat{\textbf{p}}=-i\hbar\nabla$ and $\omega = kc$. 

Inserting this last relation in equation (\ref{eq:revised Hamiltonian}), we obtain
\begin{align}
    \hat{H}&=\sum_{j=1}^N\left(\gamma mc^2\left|\textbf{r}_j\rangle\langle\textbf{r}_j\right|+\frac{\hbar\omega_0'}{\gamma}\hat{R}_3^{\prime j}\right)+\sum_p\hbar\omega_p\hat{a}_p^\dagger\hat{a}_p\nonumber\\
    &-\sum_{j=1}^N\sum_p\left[\rule{0mm}{5mm}\left(\textbf{d}'\cdot\boldsymbol{\epsilon}_p\right)\left(\hat{R}_-^{\prime j}+\hat{R}_+^{\prime j}\right)\left(\hat{E}_p^++\hat{E}_p^-\right)\right.\nonumber\\
    &\quad\left.+\frac{\hbar \omega}{\gamma mc^2}\left(\textbf{d}'\cdot\boldsymbol{\epsilon}_p\right)\left(\hat{R}_-^{\prime j}+\hat{R}_+^{\prime j}\right)\left(\hat{E}_p^+-\hat{E}_p^-\right)\right].\label{eq:full Hamiltonian}
\end{align}

Finally, since $\hbar\omega\ll \gamma mc^2$, the last term in equation (\ref{eq:full Hamiltonian}) is negligible and hence omitted in equation (\ref{eq:interaction Hamiltonian}). 

\bibliography{apssamp}

\end{document}